\documentclass[preprintnumbers,
amsmath,
prd,nofootinbib,floatfix,11pt,
]{revtex4}

\newcommand\beq{\begin{eqnarray}}
\newcommand\eeq{\end{eqnarray}}

\def\masterIfourtwo{I_{42}}
\def\masterIfourzero{I_{40}}
\def\masterIfiveone{I_{51}}
\def\masterIfivethree{I_{53}}
\def\masterIsixzero{I_{60}}

\def\masterIsixtwo{I_{62}}
\def\masterIsixthree{I_{63}}
\def\masterIsixfour{I_{64}}
\def\masterIseventwo{I_{72}}
\def\masterIseventhree{I_{73}}
\def\masterIeightfour{I_{84}}
\def\masterIninethree{I_{93}}

\def\lnbar{\overline{\ln}}
\def\lsim{\mathrel{\rlap{\lower4pt\hbox{$\sim$}}
    \raise1pt\hbox{$<$}}}                
\def\gsim{\mathrel{\rlap{\lower4pt\hbox{$\sim$}}
    \raise1pt\hbox{$>$}}}            

\allowdisplaybreaks
\def\MSbar{\overline{\rm MS}}
\usepackage{graphicx}
\usepackage{setspace}

\begin{document}
\renewcommand{\theequation}{\arabic{section}.\arabic{equation}}
\renewcommand{\thefigure}{\arabic{section}.\arabic{figure}}
\renewcommand{\thetable}{\arabic{section}.\arabic{table}}

\title{\large \baselineskip=16pt 
Four-loop Standard Model effective potential at leading order in QCD}

\author{Stephen P.~Martin}
\affiliation{
\mbox{\it Department of Physics, Northern Illinois University, DeKalb IL 60115}, \\
\mbox{\it Fermi National Accelerator Laboratory, P.O. Box 500, Batavia IL 60510}}

\begin{abstract}\normalsize \baselineskip=14pt 
The leading QCD part of the four-loop contribution to the 
effective potential for the Standard Model Higgs field is 
found. As a byproduct, I also find the corresponding 
contribution to the four-loop beta function of the Higgs 
self-interaction coupling.
\end{abstract}

\maketitle
\tableofcontents

\baselineskip=15.4pt

\section{Introduction\label{sec:intro}}
\setcounter{equation}{0}
\setcounter{figure}{0}
\setcounter{table}{0}
\setcounter{footnote}{1}

The effective potential \cite{Coleman:1973jx,Jackiw:1974cv,Sher:1988mj} 
is an important tool for analyzing spontaneous symmetry breaking 
associated with scalar field vacuum expectation values (VEVs). In the Standard 
Model, it provides a quantitative link between the Lagrangian 
parameters and the VEV of the Higgs field. The fact that the Higgs boson 
mass is near 125 GeV implies that the electroweak vacuum is close to 
metastable, motivating a program of precise study of the stability
criteria 
\cite{Sher:1988mj,Lindner:1988ww,Arnold:1991cv,Ford:1992mv,Casas:1994qy,
Espinosa:1995se,Casas:1996aq,Isidori:2001bm,Espinosa:2007qp,
ArkaniHamed:2008ym,Bezrukov:2009db,Ellis:2009tp, 
EliasMiro:2011aa,Alekhin:2012py,Bezrukov:2012sa,Degrassi:2012ry,
Buttazzo:2013uya,Jegerlehner:2013dpa,Bednyakov:2013cpa}.
Of more general importance is the fact that 
the effective potential minimization condition allows one to
determine and eliminate 
one of the Lagrangian parameters of the theory, typically the
negative Higgs squared mass parameter, in favor of the radiatively corrected
VEV.

The effective potential $V_{\rm eff}(\phi)$ can be obtained as the sum of 
one-particle irreducible vacuum Feynman graphs, computed in terms 
of particle masses and couplings that depend on a constant
background scalar field $\phi$.
In the normalization conventions of the present paper, 
the canonically normalized
Standard Model Higgs complex doublet field $\Phi$ has a tree-level potential
\beq
V &=& m^2 \Phi^\dagger \Phi +\lambda (\Phi^\dagger \Phi)^2 ,
\eeq
where $\lambda$ is the Higgs self-interaction coupling,
and the negative Higgs squared mass parameter is $m^2$.
The real neutral part of $\Phi$ is given by 
$(\phi + h)/\sqrt{2}$, where $\phi$ is the constant background field
and $h$ is the physical Higgs real scalar boson field.
The complete set of 1-loop and 2-loop 
contributions to the effective potential in Landau gauge 
are known for the Standard Model \cite{Ford:1992pn} and 
for a general renormalizable field theory \cite{Martin:2001vx}. Also 
known \cite{Martin:2013gka} are the 3-loop contributions that only 
involve the strong coupling $g_3$ and the top-quark Yukawa coupling 
$y_t$. Contributions from Goldstone bosons can be resummed 
\cite{Martin:2014bca,Elias-Miro:2014pca} in order to avoid potential 
infrared singularities and spurious imaginary parts. The value of
the background field at the minimum of the effective potential is
the radiatively corrected VEV of the Higgs field. 

The purpose of this paper is to extend the existing calculations 
of the effective potential $V_{\rm eff}(\phi)$ by 
obtaining the 4-loop contributions that are 
leading in the strong coupling $g_3$, using dimensional regularization
\cite{Bollini:1972ui,Ashmore:1972uj,Cicuta:1972jf,tHooft:1972fi,tHooft:1973mm} 
and the $\MSbar$ renormalization scheme \cite{Bardeen:1978yd,Braaten:1981dv}.
These contributions come from those
diagrams that involve only quarks, gluons, and QCD ghost fields. I will 
work in the approximation that all quarks are massless except the top 
quark. This is an excellent approximation beyond 1-loop order,
due to the small magnitudes of 
the Yukawa couplings of the bottom and other quarks. Then, in dimensional 
regularization, at least one top-quark loop must be present in a diagram 
in order for 
the contribution not to vanish. At loop order $\ell$, the resulting leading 
QCD contribution is proportional to $g_3^{2(\ell-1)} t^2$  multiplied
by a polynomial of order $\ell$ in $\lnbar(t)$, where 
\beq
t &\equiv& y_t^2 \phi^2/2,
\label{eq:deft}
\eeq
is the field-dependent $\MSbar$ top-quark squared mass, and
\beq
\lnbar(t) &\equiv& \ln(t/Q^2),
\label{eq:deflnbart}
\eeq
where $Q$ is the $\MSbar$ renormalization scale. 

The organization of the remainder of this paper is as follows. In 
section \ref{sec:basis}, I review the basis of scalar integrals used
in the calculation. 
The effective potential in 
\beq
d = 4 - 2 \epsilon,
\eeq
spacetime dimensions is given in section
\ref{sec:bare} in 
terms of bare quantities and the basis integrals. 
In section \ref{sec:renormalized}, the bare parameters 
are re-expressed in terms of $\MSbar$ quantities to obtain the effective 
potential in that renormalization scheme, after expanding in $\epsilon$. 
(This is more efficient than doing a separate 
calculation of counterterm diagrams.) In the process, I obtain the 
leading QCD contribution to the 
4-loop beta function for $\lambda$, from the requirement that poles in 
$\epsilon$ do not appear in the effective potential when written in terms of 
the renormalized parameters. Some of the results, when given in 
general form in terms of group theory invariants, 
are rather lengthy and therefore
are provided in ancillary electronic files rather than in print.
Section \ref{sec:discussion} concludes with some brief comments on the
numerical impact of the new results.

\section{Three-loop and four-loop integral basis\label{sec:basis}}
\setcounter{equation}{0}
\setcounter{figure}{0}
\setcounter{table}{0}
\setcounter{footnote}{1}

\begin{figure}[t]
\begin{center}
\includegraphics[width=0.95\linewidth,angle=0]{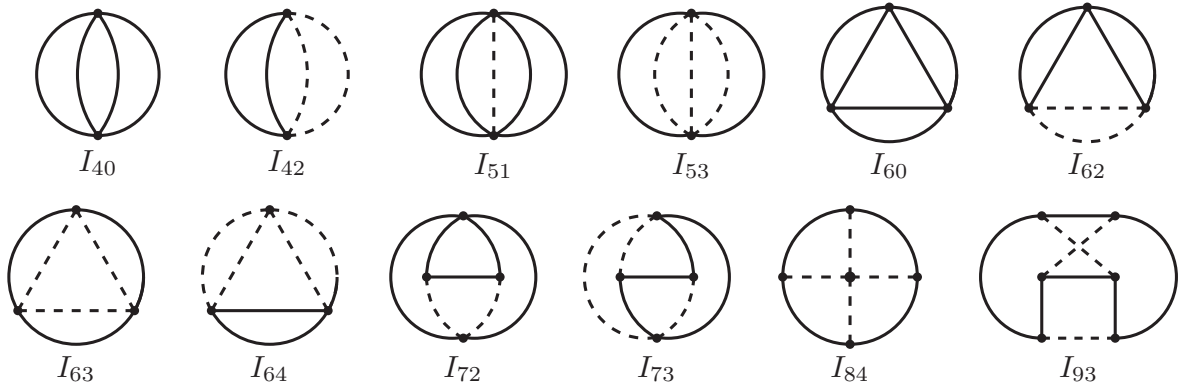}
\end{center}
\caption{\label{fig:integrals}
The 3-loop and 4-loop scalar basis integrals with one mass scale and an even number of massive propagators at each vertex. Massive propagator lines are solid, and massless propagator lines are dashed. The first subscript is the total number of
propagators, and the second is the number of massless propagators.}
\end{figure}

In the approximation of this paper, 
the only mass scale (other than the renormalization scale)
is the top-quark mass. Therefore, it is convenient to write results
in terms of Euclidean momentum integrals 
with each propagator having dimensionless mass 0 or 1. 
The dependence on the bare top-quark
mass is then restored by dimensional analysis. 
Only integrals having an even number of massive propagators 
meeting at each vertex are needed in this paper. Momentum integrations 
in $d$ dimensions are normalized by
\beq
\int_p &\equiv &\int \frac{d^dp}{(2\pi)^d},
\label{eq:intmeasure} 
\eeq
so that the 1-loop vacuum master integral is defined by
\beq
A &\equiv& \int_p \frac{1}{p^2 + 1} \>=\> 
\frac{\Gamma(1-d/2)}{(4 \pi)^{d/2}} 
.
\eeq
At 2-loop order, no new master integral appears. The necessary 3-loop and 4-loop 
integrals have been studied and used in refs.~\cite{Broadhurst:1991fi,Avdeev:1995eu,Broadhurst:1998rz,MATAD,Laporta:2002pg,
Schroder:2002re,Schroder:2005va,Schroder:2005db,Kniehl:2005yc,
Schroder:2005hy,Chetyrkin:2006dh,Chetyrkin:2006bj,Boughezal:2006xk,
Bejdakic:2006vg,Faisst:2006sr,Bejdakic:2009zz,Lee:2010hs,Liu:2015fxa}. 
Important applications include the 
calculations of the 4-loop QCD corrections 
\cite{Schroder:2005db,Chetyrkin:2006bj,Boughezal:2006xk}
to the $\rho$ parameter and 
decoupling rules for $\alpha_S$ and light quark masses across heavy quark thresholds
\cite{Schroder:2005hy,Liu:2015fxa}. 
Figure \ref{fig:integrals} shows a basis for the master integrals 
needed \cite{Schroder:2002re} for single-scale  
gauge theories at 3-loop order and 4-loop order. 
Each solid line represents a massive propagator denominator, and each dashed line
represents a massless propagator denominator, and the Euclidean
loop integrations are normalized
according to eq.~(\ref{eq:intmeasure}). 
So, for example,
\beq
\masterIfourtwo &\equiv& 
\int_p \int_q \int_k \frac{1}{p^2 q^2 (k^2 +1) [(p+q+k)^2 + 1]} .
\eeq
Also needed in the basis are products $A^2$, $A^3$, $A^4$,
$A \masterIfourzero$, and $A \masterIfourtwo$. All of the 
integrals used in this paper are reduced to the basis by repeated 
application of the integration by parts method \cite{IBP}, using a 
strategy similar to that described in ref.~\cite{Schroder:2002re}.

The integrals $\masterIfourtwo$, $\masterIfivethree$, and 
$\masterIsixfour$ have non-zero masses confined to a single 1-loop 
self-energy subdiagram, and are therefore known analytically in terms of 
$\Gamma$ functions. In general, it is sufficient to have results for the 
basis integrals as expansions in $\epsilon$. However, with the basis 
chosen here, the coefficients of the basis integrals have poles in 
$1/\epsilon$ in addition to the poles inherent in the basis 
integrals.\footnote{For an alternative basis with the nice property that 
coefficients do not contain extra poles in $\epsilon$, see 
ref.~\cite{Chetyrkin:2006dh,Faisst:2006sr}.} This means that it is 
necessary to have the expansions to certain positive powers of 
$\epsilon$ in most of the cases. The coefficients of expansions in 
$\epsilon$ of the other integrals have been given numerically with high 
precision and to sufficiently high order in $\epsilon$ in 
\cite{Schroder:2005va}, using the Laporta difference equation method 
\cite{Laporta:2001dd}. In principle this is enough for practical 
purposes, but it is nice to have analytical versions as well. These have 
been provided in 
refs.~\cite{Schroder:2005va,Bejdakic:2006vg,Bejdakic:2009zz,Lee:2010hs}. 
Table \ref{tab:basisintegrals} shows the order in $\epsilon$ to which 
each basis integral is needed in the present paper, as well as the 
highest order to which it is known analytically in terms of simple 
$\epsilon$-independent sums, and the source reference that provides that 
expansion.
\begin{table}[t!]
\caption{Summary of present analytical knowledge of 3-loop and 4-loop 
basis integrals needed in this paper and depicted in Figure \ref{fig:integrals}. 
The second row shows the order in 
the expansion in $\epsilon$ needed here. The third row 
shows the highest order in the $\epsilon$ expansion 
to which the integral is presently
known analytically in terms of simple $\epsilon$-independent sums, 
while the fourth row gives the source reference
for that expansion. All integrals were also previously
evaluated numerically to the necessary orders and beyond in ref.~\cite{Schroder:2005va}.
\label{tab:basisintegrals}}
\begin{center}
\begin{tabular}{|c||c|c|c|c|c|c|c|c|c|c|c|c|}
\hline
integral & 
  $\masterIfourzero$ & $\masterIfourtwo$ & 
  $\masterIfiveone$  & $\masterIfivethree$ & $\masterIsixzero$ & 
  $\masterIsixtwo$ & $\masterIsixthree$ & $\masterIsixfour$ & 
  $\masterIseventwo$ & $\masterIseventhree$ & 
  $\masterIeightfour$ & $\masterIninethree$
\\ \hline
needed &
  $\epsilon^3$ & $\epsilon^4$ & 
  $\epsilon^4$ & $\epsilon^3$ & $\epsilon^2$ & 
  $\epsilon^3$ & $\epsilon^2$ & $\epsilon^1$ & 
  $\epsilon^1$ & $\epsilon^1$ & 
  $\epsilon^0$ & $\epsilon^1$
\\ \hline
known & 
  $\epsilon^5$ & $\epsilon^\infty$ & 
  $\epsilon^4$ & $\epsilon^\infty$ & $\epsilon^2$ & 
  $\epsilon^3$ & $\epsilon^5$ & $\epsilon^\infty$ & 
  $\epsilon^3$ & $\epsilon^1$ & 
  $\epsilon^0$ & $\epsilon^1$
\\ \hline
source & 
  \cite{Bejdakic:2009zz}& \cite{Schroder:2005va} & 
  \cite{Lee:2010hs} & \cite{Schroder:2005va} & \cite{Lee:2010hs} & 
  \cite{Bejdakic:2009zz} & \cite{Bejdakic:2006vg} & \cite{Schroder:2005va} &
  \cite{Bejdakic:2009zz} & \cite{Lee:2010hs} &
  \cite{Lee:2010hs} & \cite{Lee:2010hs}
\\ \hline
\end{tabular}
\end{center}
\end{table}
The (probably) transcendental numbers appearing in these coefficients up 
to the orders needed in this paper are $\pi$, $\ln(2)$, and 
\beq
\zeta_n &=& \sum_{k=1}^{\infty} \frac{1}{k^n} ,
\label{eq:zetas}
\\
a_n &=& {\rm Li}_n(1/2) \>=\> \sum_{k=1}^{\infty} \frac{1}{2^k k^n},
\label{eq:as}
\\
s_6 &=& \sum_{n=1}^{\infty}\sum_{k=1}^{n} \frac{(-1)^{n+k}}{n^5 k} ,
\label{eq:ssix}
\eeq
although the last quantity cancels out of the results below. (The absence of
this quantity could presumably have been made manifest by 
using the alternative
basis of \cite{Chetyrkin:2006dh,Faisst:2006sr}.)

\section{Effective potential in terms of bare quantities\label{sec:bare}}
\setcounter{equation}{0}
\setcounter{figure}{0}
\setcounter{table}{0}
\setcounter{footnote}{1}

In this section, I find the 4-loop effective 
potential in terms of the bare quantities in $d=4 - 2\epsilon$ dimensions. 
These include the
external scalar field $\phi_B$ and the bare Yukawa coupling 
$y_{tB}$ and QCD coupling $g_{3B}$. 
In the next section, the results will be converted to $\MSbar$ parameters. 
The loop expansion for the effective potential is written as
\beq
V_{\rm eff} = \sum_{\ell=0}^\infty V^{(\ell)}_B.
\label{eq:Veffbare}
\eeq
The tree-level potential is 
\beq
V^{(0)}_B = \frac{m^2_B}{2} \phi_B^2 + \frac{\lambda_B}{4} \phi_B^4 ,
\label{eq:V0bare}
\eeq
where $\lambda_B$ and $m^2_B$ are the bare Higgs self-coupling and
squared mass parameter, respectively.
The latter will play no role in the following. 

At each loop order, the contribution to the effective potential is given by the 
sum of 1-particle irreducible Feynman diagrams with no external legs and containing only
quarks, gluons, and QCD ghosts.
The pertinent contributions at loop order $\ell \geq 1$ are proportional to 
$g_{3B}^{2\ell-2} t_B^{2 + \ell (d - 4)/2}$, where
\beq
t_B &=& y_{tB}^2 \phi_B^2/2
\eeq
is the bare field-dependent top-quark mass.
Results below will be given in terms of group theory invariants: 
the dimension of the fundamental representation
$N_c$, the Casimir invariants of the adjoint and fundamental representations 
$C_G$ and $C_F$, the Dynkin index of the fundamental representation $T_F$, 
and the number of quark flavors $n_q$. In the Standard Model, these are given by
\beq
C_G &=& N_c = 3,
\label{eq:defCG}
\\
C_F &=& \frac{N_c^2-1}{2N_c} = 4/3,
\\
T_F &=& 1/2,
\\
n_q &=& 6,
\label{eq:defNq}
\eeq
but leaving them general provides more information for comparisons and checks. 
Diagrams at 2-loop order and higher are calculated with a gluon propagator
\beq
-i [ g^{\mu\nu} / p^2 - (1-\xi) p^\mu p^\nu / (p^2)^2 ],
\eeq
where $\xi=1$ for Feynman gauge and $\xi=0$ for Landau gauge. The dependence on
the (bare) QCD gauge-fixing parameter $\xi$ cancels at the level of the 
basis integrals, providing a stringent check.

The contributions involving only quarks, gluons, and QCD ghosts, up to 3-loop order, are \cite{Martin:2013gka}:
\beq
V^{(1)}_B &=&  -4N_c t_B^{d/2} A/d ,
\label{eq:V1bare}
\\
V^{(2)}_B &=& N_c C_F g_{3B}^2 t_B^{d-2} A^2 {(d-1)(d-2)}/{(d-3)} 
,
\label{eq:V2bare}
\\
V^{(3)}_B &=& g_{3B}^4 t^{3d/2-4} N_c C_F \Bigl \{ C_G  \Bigl [
  \frac{(2-d)^3}{2 (d-4)^2 (d-3)} A^3
  + \frac{(3-d)(d^3 - 13 d^2 + 50 d - 48)}{4 (d-4)^2} \masterIfourzero
\nonumber \\ &&
  + \frac{(d-2)^2 (2 d^2 - 17 d + 32)}{2 (d-4) (2d-7)} \masterIfourtwo  
\Bigr]
+ C_F \Bigl [
  \frac{(d-6)(d-3)(d^2 - 7d + 8)}{2(d-4)^2} \masterIfourzero
\nonumber \\ &&
 + \frac{(d-2)^2 (-d^5 + 13 d^4 - 67 d^3 + 181 d^2 - 274 d + 188)}{
2 (d-4)^2 (d-3)^2} A^3
\nonumber \\ &&
 + \frac{(2-d)(2 d^3 - 21 d^2 + 67 d - 68)}{(d-4)(d-3)} \masterIfourtwo 
\Bigr ]
+ T_F \Bigl [
\frac{2(5-d)(d-2)^3}{(d-6)(d-4)(d-3)} A^3
\nonumber \\ &&
+ \frac{d^3 - 7 d^2 + 6 d + 16}{(d-6)(4-d)} \masterIfourzero 
+ (n_q-1) 
\frac{4 (d-3)(d-2)}{7 - 2 d} \masterIfourtwo
\Bigr ]
\Bigr \} 
.
\label{eq:V3bare}
\eeq

For the 4-loop order contributions involving quarks, gluons, and QCD ghosts, there are 51 Feynman diagrams, which are reduced to linear combinations of the 13 integrals from the set
\beq
{\cal I} =
\{A^4,\> A\masterIfourzero,\> A\masterIfourtwo, \> \masterIfiveone,\>
\masterIfivethree,\> \masterIsixzero,\>
\masterIsixtwo,\>
\masterIsixthree,\>
\masterIsixfour,\>
\masterIseventwo,\>   
\masterIseventhree,\>
\masterIeightfour,\>
\masterIninethree\},
\label{eq:masters}
\eeq
using integration by parts identities.
The four-loop effective potential contribution is then organized in terms of the group
theory invariants from the set
\beq
{\cal{G}} = \{C_G^2,\> C_G T_F,\> C_G T_F n_q,\> C_G C_F,\> C_F^2,
\> C_F T_F,\> C_F T_F n_q,\> T_F^2,\> T_F^2 n_q,\> T_F^2 n_q^2\} 
,
\label{eq:groupinvariants}
\eeq
so that the result is written as:  
\beq
V_B^{(4)} = g_{3B}^6 t_B^{2d-6} C_F N_c 
\sum_{\cal{G}}\sum_{{\cal I}} \, {\cal{G}}\, {\cal I}
\,
V^{(4)}_{B}({\cal{G}},{\cal{I}})
.
\label{eq:V4bare}
\eeq
The 130 coefficients $V^{(4)}_{B}({\cal{G}},{\cal{I}})$ are rational 
functions of the spacetime dimension $d$. Although 58 of them vanish,
this list of coefficients is still rather lengthy, so they are not shown in print here. 
Instead, they are provided in an ancillary electronic 
file called {\tt V4bare.txt} 
included with the arXiv submission for this article.

\section{Effective potential in terms of renormalized 
quantities\label{sec:renormalized}}
\setcounter{equation}{0}
\setcounter{figure}{0}
\setcounter{table}{0}
\setcounter{footnote}{1}

In this section, I obtain the 
effective potential in the $\MSbar$ renormalization scheme by
translating the bare quantities into
$\MSbar$ quantities. Because $\int d^dx V$ must be dimensionless in order to
be exponentiated in the path integral, one must introduce an arbitrary
regularization scale $\mu$, which is related to the $\MSbar$ renormalization scale
$Q$ by \cite{Bardeen:1978yd,Braaten:1981dv}:
\beq
Q^2 = 4\pi e^{-\gamma_E} \mu^2.
\eeq
Then, in the $\MSbar$ scheme, one writes:
\beq
\phi_B &=&  \mu^{-\epsilon} \phi \sqrt{Z_\phi} ,
\label{eq:defphiB}
\\
Z_\phi &=& 
1 + \sum_{\ell=1}^\infty \sum_{n=1}^\ell 
\frac{z^\phi_{\ell,n}}{(16\pi^2)^\ell\epsilon^n} ,
\label{eq:defZphi}
\\
x_{kB} &=&  \mu^{\rho_{x_k} \epsilon} \Bigl ( x_k + \sum_{\ell=1}^\infty \sum_{n=1}^\ell 
\frac{z^{x_k}_{\ell,n}}{(16\pi^2)^\ell \epsilon^n}
\Bigr ) .
\label{eq:defzkB}
\eeq
The subscript $B$ labels bare quantities, while the absence of
a subscript $B$ indicates the corresponding $\MSbar$ renormalized quantity. 
The exponent $\ell$ is the loop order, while 
$k$ is an index that runs over the list of Lagrangian parameters, including
$x_k = \lambda, y_t, g_3$. The mass dimensions of the 
bare parameters determine that $\rho_\lambda = 2$ and $\rho_{g_3} = \rho_{y_t} = 1$, in
order that the renormalized couplings $\lambda, g_3$, and $y_t$ are dimensionless
and $\phi$ has mass dimension 1.
The counter-term quantities $z^{\phi}_{\ell,n}$ and $z^{x_k}_{\ell,n}$ are 
polynomials in the $\MSbar$ renormalized parameters $x_j$, and 
do not depend on $\epsilon$ or $\phi$. They are determined by the 
requirement that the full effective potential (and all physical 
observables) are free of ultraviolet poles in $\epsilon$ when expressed in terms of 
the $\MSbar$ quantities.

The anomalous dimension
for $\phi$ and the $\MSbar$ beta functions for the parameters $x_k$ are defined by
\beq
\gamma &\equiv& 
-Q \frac{d \ln\phi}{dQ} \Bigl |_{\epsilon=0}
\>=\> 
-Q \frac{d \ln\phi}{dQ} + \epsilon
\>=\> 
\frac{1}{2} Q \frac{d}{dQ} \ln(Z_\phi) 
,
\\
\beta_{x_k} &\equiv&
Q \frac{d x_k}{dQ} \Bigl |_{\epsilon=0}
\>=\>
Q \frac{d x_k}{dQ} + \epsilon \rho_{x_k} x_k .
\eeq
Because the bare quantities $\phi_B$ and $x_{kB}$ 
do not depend on $Q$, the
anomalous dimension and beta functions are determined  
by the simple pole counterterms, so that:
\beq
\gamma &=& 
\sum_{\ell=1}^\infty \frac{1}{(16 \pi^2)^\ell} \gamma^{(\ell)},
\label{eq:gammaexp}
\\
\beta_{x_k} &=& 
\sum_{\ell=1}^\infty \frac{1}{(16 \pi^2)^\ell} \beta_{x_k}^{(\ell)},
\label{eq:betaxkexp}
\eeq
where the $\ell$-loop contributions are:
\beq
\gamma^{(\ell)} &=& -\ell z^\phi_{\ell,1},
\label{eq:gammaafromcphi}
\\
\beta_{x_k}^{(\ell)} &=& 2 \ell z^{x_k}_{\ell,1} .
\label{eq:betafromck}
\eeq
The higher pole counterterms are also fixed by consistency conditions 
\beq
\ell z^\phi_{\ell,n} &=&
\sum_{\ell'=1}^{\ell-n+1} \Bigl (
-\gamma^{(\ell')} + 
\frac{1}{2} \sum_j \beta_{x_j}^{(\ell')} \frac{\partial}{\partial x_j}
\Bigr ) z^\phi_{\ell-\ell',n-1}
,
\label{eq:higherpolecphi}
\\
2 \ell z^{x_k}_{\ell,n}
&=&
\sum_{\ell'=1}^{\ell-n+1}\sum_j \beta_{x_j}^{(\ell')} 
\frac{\partial}{\partial x_j}
z^{x_k}_{\ell-\ell',n-1}
.
\label{eq:higherpoleck}
\eeq
for $\ell \geq n \geq 2$.

The coefficients $z^\phi_{\ell,n}$ and $z^{x_k}_{\ell,n}$ for $\ell \leq 3$ 
are thus determined by the known results for the Standard Model beta functions and Higgs
scalar anomalous dimension given in
\cite{MVI,MVII,Jack:1984vj,MVIII,Chetyrkin:2012rz,Chetyrkin:2013wya,Bednyakov:2013eba}. 
(Extensions to QCD 4-loop and 5-loop order can be found in 
\cite{vanRitbergen:1997va,Czakon:2004bu,Chetyrkin:1997dh,
Vermaseren:1997fq,Baikov:2014qja,Marquard:2015qpa}.)
Keeping only the contributions
needed for the approximation of the present paper, they are:
\beq
z^{\lambda}_{1,1} &=& 
-N_c y_t^4 + \ldots ,
\\
z^{\lambda}_{2,1} &=& 
g_3^2 y_t^4 (-2 N_c C_F)  
+ \ldots ,
\\
z^{\lambda}_{2,2} &=& g_3^2 y_t^4 (6 N_c C_F)  
+ \ldots ,
\\
z^{\lambda}_{3,1} &=& g_3^4 y_t^4 N_c C_F 
\left [ \left (8 \zeta_3 -\frac{109}{6} \right ) C_G 
+ \left (\frac{131}{6} - 16 \zeta_3\right ) C_F 
+ \left (16 + \frac{10}{3} n_q \right ) T_F  
\right ]
+ \ldots ,
\\
z^{\lambda}_{3,2} &=& g_3^4 y_t^4  N_c C_F \left (
24 C_G + 10 C_F - 16 T_F n_q/3  
\right ) + \ldots ,
\\
z^{\lambda}_{3,3} &=& g_3^4 y_t^4 N_c C_F   \left (
-22 C_G/3 -24 C_F + 8 T_F n_q/3  
\right )
+ \ldots ,
\\
z^{y_t}_{1,1} &=& g_3^2 y_t (-3 C_F) + \ldots 
,
\\
z^{y_t}_{2,1} &=& g_3^4 y_t C_F \Bigl ( -\frac{97}{12} C_G - \frac{3}{4} C_F +
\frac{5}{3} T_F n_q \Bigr ) 
+ \ldots ,
\phantom{xxx}
\\
z^{y_t}_{2,2} &=& g_3^4 y_t C_F \Bigl ( \frac{11}{2} C_G 
+ \frac{9}{2} C_F - 2 T_F n_q \Bigr )  
+ \ldots ,
\\
z^{y_t}_{3,1} &=& g_3^6 y_t C_F \biggl [ 
-\frac{11413}{324} C_G^2 
+ \frac{43}{4} C_G C_F 
- \frac{43}{2} C_F^2 
+ \left (\frac{556}{81} + 16 \zeta_3 \right ) C_G T_F n_q 
\nonumber \\ &&
+ \left (\frac{46}{3} - 16 \zeta_3 \right ) C_F T_F n_q
+ \frac{140}{81} T_F^2 n_q^2
\biggr )  
+ \ldots ,
\\
z^{y_t}_{3,2} &=& g_3^6 y_t C_F \Bigl ( 
\frac{1679}{54} C_G^2 
+ \frac{313}{12} C_F C_G 
+ \frac{9}{4} C_F^2 
-\frac{484}{27} C_G T_F n_q
\nonumber \\ &&
-\frac{29}{3} C_F T_F n_q
+\frac{40}{27} T_F^2 n_q^2
\Bigr )  
+ \ldots ,
\\
z^{y_t}_{3,3} &=& g_3^6 y_t C_F \Bigl ( 
-\frac{121}{9} C_G^2 + \frac{88}{9} C_G T_F n_q
- \frac{33}{2} C_G C_F - \frac{9}{2} C_F^2 
+ 6 C_F T_F n_q 
- \frac{16}{9} T_F^2 n_q^2 \Bigr )  
+ \ldots,\phantom{xxxx}
\\
z^{g_3}_{1,1} &=& g_3^3 \Bigl (
-\frac{11}{6} C_G + \frac{2}{3} T_F n_q
\Bigr ) ,
\\
z^{g_3}_{2,1} &=& g_3^5 \Bigl (
-\frac{17}{6} C_G^2 + \frac{5}{3} C_G T_F n_q + C_F T_F n_q 
\Bigr ) + \ldots ,
\\
z^{g_3}_{2,2} &=& g_3^5 \Bigl (
\frac{121}{24} C_G^2 - \frac{11}{3} C_G T_F n_q + \frac{2}{3} T_F^2 n_q^2 
\Bigr ) + \ldots ,
\eeq
while the $z^{\phi}_{\ell,n}$ do not contribute at all at leading order in QCD.
Now, expanding eq.~(\ref{eq:Veffbare}) with eqs.~(\ref{eq:V0bare}), 
(\ref{eq:V1bare}), (\ref{eq:V2bare}), (\ref{eq:V3bare}), and (\ref{eq:V4bare}) to order $1/\epsilon$, and requiring the 4-loop simple pole terms to cancel, I find:
\beq
z^{\lambda}_{4,1} &=& y_t^4 g_3^6 N_c C_F \Biggl [
C_G^2 \left( \frac{470}{3} \zeta_3 -130 \zeta_5 -\frac{121547}{972} 
   -\frac{11 \pi^4}{45}\right)
+ C_G T_F \left(\frac{1472}{9} + 88 \zeta_3 -40 \zeta_5 \right)
\nonumber \\ &&
+ C_G T_F n_q \left(4 \zeta_3 -\frac{661}{243} +\frac{16 \pi^4}{45} \right)
+ C_G C_F \left(\frac{896}{3}-\frac{826}{3} \zeta_3 + 180 \zeta_5 +  
  \frac{4 \pi^4}{45} \right)
\nonumber \\ &&
+C_F^2 \left(12 \zeta_3 + 40 \zeta_5 -\frac{1471}{6} + \frac{4 \pi^4}{5}\right)
-16 C_F T_F
+ C_F T_F n_q \left(\frac{281}{6} + \frac{8}{3} \zeta_3 - \frac{4 \pi^4}{9}\right)
\nonumber \\ &&
- T_F^2 n_q \frac{64}{9} 
+ T_F^2 n_q^2 \left(\frac{2728}{243}-\frac{32}{3}\zeta_3 \right)
\Biggr ] + \ldots,
\label{eq:zlambda41}
\\
z^{\lambda}_{4,2} &=& y_t^4 g_3^6 N_c C_F \biggl [
\left (\frac{7811}{54}-\frac{44 \zeta_3}{3}\right ) C_G^2 
-\frac{88}{3} C_G T_F 
+ \left (-\frac{128 \zeta_3}{3}-\frac{1138}{27} \right ) C_G T_F n_q 
\nonumber \\ &&
+ \left (\frac{131}{9} + \frac{16}{3} \zeta_3 \right ) C_G C_F 
+ \left (2 + 48 \zeta_3 \right ) C_F^2
- 48 C_F T_F 
\nonumber \\ &&
+ \left (-\frac{451}{9} + \frac{112}{3} \zeta_3 \right ) C_F T_F n_q 
+ \frac{32}{3} T_F^2 n_q 
-\frac{80}{27} T_F^2 n_q^2
\biggr ] + \ldots,
\\
z^{\lambda}_{4,3} &=& y_t^4 g_3^6 N_c C_F \biggl [
-61 C_G^2 +\frac{322}{9} C_G T_F n_q -\frac{562}{3} C_G C_F - 39 C_F^2 + 
\frac{146}{3} C_F T_F n_q 
\nonumber \\ &&
-\frac{32}{9} T_F^2 n_q^2
\biggr ] + \ldots,
\\
z^{\lambda}_{4,4} &=& y_t^4 g_3^6 N_c C_F \biggl [
\frac{121}{9} C_G^2 -\frac{88}{9} C_G T_F n_q + 66 C_G C_F + 72 C_F^2 - 24 C_F T_F n_q 
\nonumber \\ &&
+ \frac{16}{9} T_F^2 n_q^2
\biggr ] + \ldots,
\eeq
where the ellipses refer to contributions that are lower order in $g_3$.
From eqs~(\ref{eq:betafromck}) and (\ref{eq:zlambda41}), 
I find the leading QCD 4-loop contribution to $\beta_\lambda$:
\beq
\beta^{(4)}_\lambda &=& y_t^4 g_3^6 N_c C_F \Biggl [
C_G^2 \left( \frac{3760}{3} \zeta_3 -1040 \zeta_5 -\frac{243094}{243} 
   -\frac{88 \pi^4}{45}\right)
+ C_G T_F \left(\frac{11776}{9} + 704 \zeta_3 -320 \zeta_5 \right)
\nonumber \\ &&
+ C_G T_F n_q \left(32 \zeta_3 -\frac{5288}{243} +\frac{128 \pi^4}{45} \right)
+ C_G C_F \left(\frac{7168}{3}-\frac{6608}{3} \zeta_3 + 1440 \zeta_5 +  
  \frac{32 \pi^4}{45} \right)
\nonumber \\ &&
+C_F^2 \left(96 \zeta_3 + 320 \zeta_5 -\frac{5884}{3} + \frac{32 \pi^4}{5}\right)
-128 C_F T_F
+ C_F T_F n_q \left(\frac{1124}{3} + \frac{64}{3} \zeta_3 - \frac{32 \pi ^4}{9}\right)
\nonumber \\ &&
- T_F^2 n_q \frac{512}{9} 
+ T_F^2 n_q^2 \left(\frac{21824}{243}-\frac{256}{3}\zeta_3 \right)
\Biggr ]   + \ldots .
\label{eq:beta4lambda}
\eeq
Now taking the limit $\epsilon \rightarrow 0$,
the effective potential is obtained in a loop expansion as
\beq
V_{\rm eff} = \sum_{\ell= 0}^\infty \frac{1}{(16\pi^2)^\ell} V^{(\ell)}.
\eeq
Note that unlike the loop expansion with bare parameters, eq.~(\ref{eq:Veffbare}), here
loop factors $1/(16 \pi^2)^\ell$ have been extracted, 
similarly to eqs.~(\ref{eq:gammaexp}) and (\ref{eq:betaxkexp}).
In terms of $t$ and $\lnbar(t)$ defined in 
eqs.~(\ref{eq:deft}) and (\ref{eq:deflnbart}),
the previously known 
results for the leading QCD effective potential contributions are:
\beq
V^{(0)} &=& \frac{m^2}{2} \phi^2 + \frac{\lambda}{4} \phi^4 ,
\label{eq:V0ren}
\\
V^{(1)} &=& -N_c t^2 \bigl [\lnbar(t) - 3/2 \bigr ]
,
\label{eq:V1ren}
\\
V^{(2)} &=& g_3^2 N_c C_F t^2 \bigl [6 \lnbar^2(t) - 16 \lnbar(t) + 18 \bigr ]
,\phantom{xxx}
\label{eq:V2ren}
\eeq
from ref.~\cite{Ford:1992pn}, and the three-loop result \cite{Martin:2013gka}:
\beq
V^{(3)} &=& g_3^4 N_c C_F t^2 \Bigl \{
C_G \Bigr [-\frac{22}{3} \lnbar^3(t) 
             + \frac{185}{3} \lnbar^2(t)
             + (24 \zeta_3 - \frac{1111}{6}) \lnbar(t)
\nonumber \\ &&
             + \frac{2609}{12} 
             + \frac{44}{45}\pi^4 
             - \frac{232}{3} \zeta_3 
             + \frac{16}{3} \ln^2(2) [\pi^2 - \ln^2(2)]
             - 128 a_4 
\Bigl ]
\nonumber \\ &&
+
C_F \Bigr [-24 \lnbar^3(t) 
             + 63 \lnbar^2(t)
             - (48 \zeta_3 + \frac{121}{2}) \lnbar(t)
             + \frac{85}{12} 
             - \frac{88}{45}\pi^4 
\nonumber \\ &&
             + 192 \zeta_3 
             - \frac{32}{3} \ln^2(2) [\pi^2 - \ln^2(2)]
             + 256 a_4 
\Bigl ]
+
T_F \Bigr [48 \lnbar(t)
             - \frac{232}{3} 
             + 96 \zeta_3 
\Bigl ]
\nonumber \\ &&
+
T_F n_q \Bigr [\frac{8}{3} \lnbar^3(t) 
             - \frac{52}{3} \lnbar^2(t)
              + \frac{142}{3} \lnbar(t)
             - \frac{161}{3} 
             - \frac{64}{3} \zeta_3 
\Bigl ] 
\Bigr \}.
\label{eq:V3ren}
\eeq
The new 4-loop result (with group-theory quantities left general) takes the form:
\beq
V^{(4)} = g_{3}^6 C_F N_c t^2 
\sum_{\cal{G}}\sum_{n=0}^{4} \, {\cal{G}}\, \lnbar^n(t)
\,
V^{(4)}({\cal{G}},n),
\eeq
in terms of the group theory invariants in the set 
${\cal G}$ from eq.~(\ref{eq:groupinvariants}).
The list of 50 coefficients $V^{(4)}({\cal{G}},n)$ is again rather 
lengthy, and so is provided in another ancillary electronic
file {\tt V4MSbar.txt}. 
After substituting in the Standard Model values for the group theory 
constants, the result combines and simplifies to:
\beq
V^{(4)} &=& g_3^6 t^2 \Biggl [
\frac{13820381}{270} + \frac{1747112 \zeta_3}{45}
+\frac{1984 \zeta_5}{9} -\frac{40288 \zeta_3^2}{9} 
-\frac{298894 \pi^4}{1215} 
-\frac{1780 \pi^6}{243} 
+\frac{5888 \ln^5(2)}{135}
\nonumber \\ &&
-\frac{5888}{81} \pi^2 \ln^3(2) 
-\frac{36064}{405} \pi^4 \ln(2)
+\frac{78464}{81} \ln^2(2)[\ln^2(2) - \pi^2]  
+\frac{627712 a_4}{27} -\frac{47104 a_5}{9} 
\nonumber \\ &&
+\lnbar(t) \left(\frac{27680 \zeta_3}{3}-\frac{63200 \zeta_5}{9}
-\frac{1547146}{27}-\frac{208 \pi^4}{9}
+\frac{640}{3}\ln^2(2)[\ln^2(2) - \pi^2]  + 5120 a_4\right)
\nonumber \\ &&
+ (30584-2400 \zeta_3) \lnbar^2(t) 
- 9144 \lnbar^3(t)
+ 1380 \lnbar^4(t) 
\Biggr ] .
\label{eq:V4MS}
\eeq
Equation (\ref{eq:V4MS}) can be consistently 
added to the 3-loop effective potential
as given in refs.~\cite{Ford:1992pn} and \cite{Martin:2013gka}. Also,
the condition for the minimum $v = \phi_{\rm min}$ of the Landau
gauge effective potential of the Standard Model (including
the effects of resummation of the Goldstone boson contributions from
the terms up to 3-loop order) is obtained by subtracting 
\beq
\frac{1}{(16 \pi^2)^4} \widehat\Delta_4 &=& \frac{1}{(16 \pi^2)^4}
\frac{1}{v} \left.\frac{\partial V^{(4)}}{\partial \phi}\right |_{\phi = v}
\eeq
computed using eq.~(\ref{eq:deft}) and (\ref{eq:deflnbart}) above, 
from the right-hand side of eq.~(4.18) of ref.~\cite{Martin:2014bca}.

\section{Discussion\label{sec:discussion}}
\setcounter{equation}{0}
\setcounter{figure}{0}
\setcounter{table}{0}
\setcounter{footnote}{1}

The main results of this paper are the leading QCD 4-loop contributions 
to the Higgs self-coupling beta function $\beta_\lambda$ and to the 
effective potential and its minimization condition. In each case, it is 
certainly possible that other contributions at 4-loop order, and the 
presently unknown 3-loop effects involving electroweak couplings in the 
case of the effective potential, could be numerically comparable to or 
even larger than the ones found here. The same is certainly
true of parametric 
uncertainties from the top-quark Yukawa coupling (or mass) and the 
strong coupling. Therefore the results found here are perhaps most 
useful, for the present, as ways of formalizing estimates of purely 
theoretical error.

The 4-loop leading QCD contribution of eq.~(\ref{eq:beta4lambda}) to the $\lambda$ beta
function can be expressed in numerical form as 
\beq
\Delta \beta_\lambda &=&  \, \frac{1}{(16 \pi^2)^4} \> 8308.17 g_3^6 y_t^4 .
\eeq
This can be compared to the leading QCD 1, 2, and 3-loop contributions:
\beq
\beta^{\rm{leading~QCD}}_\lambda &=& 
\frac{1}{16 \pi^2} (-6 y_t^4)
+ \frac{1}{(16 \pi^2)^2} (-32 g_3^2 y_t^4)
+ \frac{1}{(16 \pi^2)^3} (-100.402 g_3^4 y_t^4).
\label{eq:beta3loopLOQCD}
\eeq
We see that the 4-loop contribution has a sign opposite to that
of the other terms, and is larger in magnitude than
one might have expected from a simple geometric progression. 
However, its magnitude is still only half as big as the 
3-loop term in eq.~(\ref{eq:beta3loopLOQCD}) even at $Q=M_t$, 
and in absolute terms 
it makes only a tiny difference in extrapolating $\lambda$ to high energy scales.

The effective potential contribution of eq.~(\ref{eq:V4MS}) can similarly be expressed 
in numerical form as:
\beq
V^{(4)} &=& g_3^6 t^2 \Bigl [59366.97 - 54056.36 \, \lnbar(t) + 
  27699.06 \, \lnbar^2(t) - 9144\, \lnbar^3(t) + 1380\, \lnbar^4(t) \Bigr ] .
\eeq
It follows that 
the corresponding contribution to the effective potential
minimization condition 
\beq
m^2 + \lambda v^2 &=& 
-\frac{1}{16 \pi^2} \widehat \Delta_1
-\frac{1}{(16 \pi^2)^2} \widehat \Delta_2
-\frac{1}{(16 \pi^2)^3} \widehat \Delta_3
-\frac{1}{(16 \pi^2)^4} \widehat \Delta_4 + \ldots
\label{eq:mincon}
\eeq
is, numerically:
\beq
\widehat \Delta_4
&=&
g_3^6 y_t^2 t \left [
64677.58 - 52714.59\, \lnbar(t) + 27966.13\, \lnbar^2 (t)
- 12768\, \lnbar^3(t) 
+ 2760\, \lnbar^4 (t) \right] ,
\label{eq:Delta4}
\eeq
where $\widehat \Delta_\ell$ for $\ell=1,2,3$ were given 
in ref.~\cite{Martin:2014bca}.
Consider the VEV and other $\MSbar$ parameters of the 
Standard Model at benchmark values 
\beq
v(M_t) &=& \mbox{246.647 GeV},
  \label{eq:inputvev}
\\
\lambda(M_t) &=& 0.12597,
  \label{eq:inputlambda}
\\
y_t(M_t) &=& 0.93690,
  \label{eq:inputyt}
\\
g_3(M_t) &=& 1.1666,
  \label{eq:inputg3}
\\
g(M_t) &=& 0.647550,
  \label{eq:inputg}
\\
g'(M_t) &=& 0.358521,
  \label{eq:inputgp}
\eeq
at $Q = M_t = 173.34$ GeV.
These choices provide agreement with
the measured values of the $h$, $W$, and $Z$ boson masses 
in the pure $\MSbar$ scheme 
\cite{Martin:2014cxa,Martin:2015lxa,Martin:2015rea}.
Using only the previously known 3-loop contributions in eq.~(\ref{eq:mincon}),
the resulting Higgs squared mass parameter is:
$m^2(M_t) = -(92.890\>{\rm GeV})^2$.
Now including the new contribution of eq.~(\ref{eq:Delta4}) gives instead
$m^2(M_t) = -(92.926\>{\rm GeV})^2$. Thus I find
\beq
\Delta\left (\sqrt{-m^2} \right ) &=& 36\>{\rm MeV}
\eeq
from the leading QCD 4-loop contribution, at the scale $Q=M_t$. 
The parameter $m^2$ is not directly constrained by 
experiment, but it can be connected to ultraviolet completions that may 
predict it in terms of other underlying parameters that can be measured, 
eventually. This could occur in models of supersymmetry breaking, for 
example.

{\it Acknowledgments:} This work was supported in part by the National 
Science Foundation grant number PHY-1417028.



\begin{thebibliography}{90}
\baselineskip=15.2pt

\bibitem{Coleman:1973jx} 
  S.~R.~Coleman and E.~J.~Weinberg,
  ``Radiative Corrections as the Origin of Spontaneous Symmetry Breaking,''
  Phys.\ Rev.\ D {\bf 7}, 1888 (1973).

\bibitem{Jackiw:1974cv} 
  R.~Jackiw,
  ``Functional evaluation of the effective potential,''
  Phys.\ Rev.\ D {\bf 9}, 1686 (1974).

\bibitem{Sher:1988mj} 
  M.~Sher,
  ``Electroweak Higgs Potentials and Vacuum Stability,''
  Phys.\ Rept.\  {\bf 179}, 273 (1989),
  and references therein.

\bibitem{Lindner:1988ww} 
  M.~Lindner, M.~Sher and H.~W.~Zaglauer,
  ``Probing Vacuum Stability Bounds at the Fermilab Collider,''
  Phys.\ Lett.\ B {\bf 228}, 139 (1989).

\bibitem{Arnold:1991cv} 
  P.~B.~Arnold and S.~Vokos,
  ``Instability of hot electroweak theory: bounds on m(H) and M(t),''
  Phys.\ Rev.\ D {\bf 44}, 3620 (1991).

\bibitem{Ford:1992mv} 
  C.~Ford, D.~R.~T.~Jones, P.~W.~Stephenson and M.~B.~Einhorn,
  ``The Effective potential and the renormalization group,''
  Nucl.\ Phys.\ B {\bf 395}, 17 (1993)
  [hep-lat/9210033].

\bibitem{Casas:1994qy} 
  J.~A.~Casas, J.~R.~Espinosa and M.~Quir\'os,
  ``Improved Higgs mass stability bound in the standard model and implications for supersymmetry,''
  Phys.\ Lett.\ B {\bf 342}, 171 (1995)
  [hep-ph/9409458].

\bibitem{Espinosa:1995se} 
  J.~R.~Espinosa and M.~Quiros,
  ``Improved metastability bounds on the standard model Higgs mass,''
  Phys.\ Lett.\ B {\bf 353}, 257 (1995)
  [hep-ph/9504241].

\bibitem{Casas:1996aq}
  J.~A.~Casas, J.~R.~Espinosa and M.~Quiros,
  ``Standard model stability bounds for new physics within LHC reach,''
  Phys.\ Lett.\ B {\bf 382} (1996) 374
  [hep-ph/9603227].

\bibitem{Isidori:2001bm} 
  G.~Isidori, G.~Ridolfi and A.~Strumia,
  ``On the metastability of the standard model vacuum,''
  Nucl.\ Phys.\ B {\bf 609}, 387 (2001)
  [hep-ph/0104016].

\bibitem{Espinosa:2007qp} 
  J.~R.~Espinosa, G.~F.~Giudice and A.~Riotto,
  ``Cosmological implications of the Higgs mass measurement,''
  JCAP {\bf 0805}, 002 (2008)
  [0710.2484].

\bibitem{ArkaniHamed:2008ym} 
  N.~Arkani-Hamed, S.~Dubovsky, L.~Senatore and G.~Villadoro,
  ``(No) Eternal Inflation and Precision Higgs Physics,''
  JHEP {\bf 0803}, 075 (2008)
  [0801.2399].

\bibitem{Bezrukov:2009db} 
  F.~Bezrukov and M.~Shaposhnikov,
  ``Standard Model Higgs boson mass from inflation: Two loop analysis,''
  JHEP {\bf 0907}, 089 (2009)
  [0904.1537].
  
\bibitem{Ellis:2009tp} 
  J.~Ellis, J.~R.~Espinosa, G.~F.~Giudice, A.~Hoecker and A.~Riotto,
  ``The Probable Fate of the Standard Model,''
  Phys.\ Lett.\ B {\bf 679}, 369 (2009)
  [0906.0954].

\bibitem{EliasMiro:2011aa} 
  J.~Elias-Miro, J.~R.~Espinosa, G.~F.~Giudice, G.~Isidori, A.~Riotto and A.~Strumia,
  ``Higgs mass implications on the stability of the electroweak vacuum,''
  Phys.\ Lett.\ B {\bf 709}, 222 (2012)
  [1112.3022].

\bibitem{Alekhin:2012py} 
  S.~Alekhin, A.~Djouadi and S.~Moch,
  ``The top quark and Higgs boson masses and the stability of the electroweak vacuum,''
  Phys.\ Lett.\ B {\bf 716}, 214 (2012)
  [1207.0980].

\bibitem{Bezrukov:2012sa} 
  F.~Bezrukov, M.~Y.~Kalmykov, B.~A.~Kniehl and M.~Shaposhnikov,
  ``Higgs Boson Mass and New Physics,''
  JHEP {\bf 1210}, 140 (2012)
  [1205.2893].

\bibitem{Degrassi:2012ry} 
  G.~Degrassi, S.~Di Vita, J.~Elias-Miro, J.~R.~Espinosa, G.~F.~Giudice, G.~Isidori and A.~Strumia,
  ``Higgs mass and vacuum stability in the Standard Model at NNLO,''
  JHEP {\bf 1208}, 098 (2012)
  [1205.6497].

\bibitem{Buttazzo:2013uya} 
  D.~Buttazzo, G.~Degrassi, P.~P.~Giardino, G.~F.~Giudice, F.~Sala, A.~Salvio and A.~Strumia,
  ``Investigating the near-criticality of the Higgs boson,''
  [1307.3536].

\bibitem{Jegerlehner:2013dpa} 
  F.~Jegerlehner, M.~Y.~Kalmykov and B.~A.~Kniehl,
  ``About the EW contribution to the relation between pole and MS-masses of the top-quark in the Standard Model,''
  [1307.4226].

\bibitem{Bednyakov:2013cpa} 
  A.~V.~Bednyakov, A.~F.~Pikelner and V.~N.~Velizhanin,
  ``Three-loop Higgs self-coupling beta-function in the Standard Model with complex Yukawa matrices,''
  [1310.3806].

\bibitem{Ford:1992pn} 
  C.~Ford, I.~Jack and D.R.T.~Jones,
  ``The Standard model effective potential at two loops,''
  Nucl.\ Phys.\ B {\bf 387}, 373 (1992)
  [Erratum-ibid.\ B {\bf 504}, 551 (1997)]
  [hep-ph/0111190].
See also C.~Ford and D.~R.~T.~Jones,
  ``The Effective potential and the differential equations method for Feynman integrals,''
  Phys.\ Lett.\ B {\bf 274}, 409 (1992)
  [Erratum-ibid.\ B {\bf 285}, 399 (1992)].
  
\bibitem{Martin:2001vx} 
  S.P.~Martin,
  ``Two loop effective potential for a general renormalizable theory and softly broken supersymmetry,''
  Phys.\ Rev.\ D {\bf 65}, 116003 (2002)
  [hep-ph/0111209].

\bibitem{Martin:2013gka} 
  S.~P.~Martin,
  ``Three-loop Standard Model effective potential at leading order in strong and top Yukawa couplings,''
  Phys.\ Rev.\ D {\bf 89}, no. 1, 013003 (2014)
  [1310.7553].

\bibitem{Martin:2014bca} 
  S.~P.~Martin,
  ``Taming the Goldstone contributions to the effective potential,''
  Phys.\ Rev.\ D {\bf 90}, no. 1, 016013 (2014)
  [1406.2355].

\bibitem{Elias-Miro:2014pca} 
  J.~Elias-Miro, J.~R.~Espinosa and T.~Konstandin,
  ``Taming Infrared Divergences in the Effective Potential,''
  JHEP {\bf 1408}, 034 (2014)
  [1406.2652].



\bibitem{Bollini:1972ui} 
  C.~G.~Bollini and J.~J.~Giambiagi,
  ``Dimensional Renormalization: The Number of Dimensions as a Regularizing Parameter,''
  Nuovo Cim.\ B {\bf 12}, 20 (1972).
C.~G.~Bollini and J.~J.~Giambiagi,
  ``Lowest order divergent graphs in nu-dimensional space,''
  Phys.\ Lett.\ B {\bf 40}, 566 (1972).
  
\bibitem{Ashmore:1972uj} 
  J.~F.~Ashmore,
  ``A Method of Gauge Invariant Regularization,''
  Lett.\ Nuovo Cim.\  {\bf 4}, 289 (1972).

\bibitem{Cicuta:1972jf} 
  G.~M.~Cicuta and E.~Montaldi,
  ``Analytic renormalization via continuous space dimension,''
  Lett.\ Nuovo Cim.\  {\bf 4}, 329 (1972).

\bibitem{tHooft:1972fi} 
  G.~'t Hooft and M.~J.~G.~Veltman,
  ``Regularization and Renormalization of Gauge Fields,''
  Nucl.\ Phys.\ B {\bf 44}, 189 (1972).

\bibitem{tHooft:1973mm} 
  G.~'t Hooft,
  ``Dimensional regularization and the renormalization group,''
  Nucl.\ Phys.\ B {\bf 61}, 455 (1973).



\bibitem{Bardeen:1978yd} 
  W.~A.~Bardeen, A.~J.~Buras, D.~W.~Duke and T.~Muta,
  ``Deep Inelastic Scattering Beyond the Leading Order in Asymptotically Free Gauge Theories,''
  Phys.\ Rev.\ D {\bf 18}, 3998 (1978).
  
\bibitem{Braaten:1981dv} 
  E.~Braaten and J.~P.~Leveille,
  ``Minimal Subtraction and Momentum Subtraction in {QCD} at Two Loop Order,''
  Phys.\ Rev.\ D {\bf 24}, 1369 (1981).


\bibitem{Broadhurst:1991fi} 
  D.~J.~Broadhurst,
  ``Three loop on-shell charge renormalization without integration: 
  Lambda-MS (QED) to four loops,''
  Z.\ Phys.\ C {\bf 54}, 599 (1992).

\bibitem{Avdeev:1995eu} 
  L.~V.~Avdeev,
  ``Recurrence relations for three loop prototypes of bubble diagrams with a mass,''
  Comput.\ Phys.\ Commun.\  {\bf 98}, 15 (1996)
  [hep-ph/9512442].

\bibitem{Broadhurst:1998rz} 
  D.~J.~Broadhurst,
  ``Massive three-loop Feynman diagrams reducible to SC* primitives of algebras of the sixth root of unity,''
  Eur.\ Phys.\ J.\ C {\bf 8}, 311 (1999)
  [hep-th/9803091].

\bibitem{MATAD} M.~Steinhauser,
  ``MATAD: A Program package for the computation of MAssive TADpoles,''
  Comput.\ Phys.\ Commun.\  {\bf 134}, 335 (2001)
  [hep-ph/0009029].
  
\bibitem{Laporta:2002pg} 
  S.~Laporta,
  ``High precision epsilon expansions of massive four loop vacuum bubbles,''
  Phys.\ Lett.\ B {\bf 549}, 115 (2002)
  [hep-ph/0210336].

\bibitem{Schroder:2002re} 
  Y.~Schroder,
  ``Automatic reduction of four loop bubbles,''
  Nucl.\ Phys.\ Proc.\ Suppl.\  {\bf 116}, 402 (2003)
  [hep-ph/0211288].

\bibitem{Schroder:2005va} 
  Y.~Schroder and A.~Vuorinen,
  ``High-precision epsilon expansions of single-mass-scale four-loop vacuum bubbles,''
  JHEP {\bf 0506}, 051 (2005)
  [hep-ph/0503209].

\bibitem{Schroder:2005db} 
  Y.~Schroder and M.~Steinhauser,
  ``Four-loop singlet contribution to the rho parameter,''
  Phys.\ Lett.\ B {\bf 622}, 124 (2005)
  [hep-ph/0504055].

\bibitem{Kniehl:2005yc} 
  B.~A.~Kniehl and A.~V.~Kotikov,
  ``Calculating four-loop tadpoles with one non-zero mass,''
  Phys.\ Lett.\ B {\bf 638}, 531 (2006)
  [hep-ph/0508238].

\bibitem{Schroder:2005hy} 
  Y.~Schroder and M.~Steinhauser,
  ``Four-loop decoupling relations for the strong coupling,''
  JHEP {\bf 0601}, 051 (2006)
  [hep-ph/0512058].

\bibitem{Chetyrkin:2006dh} 
  K.~G.~Chetyrkin, M.~Faisst, C.~Sturm and M.~Tentyukov,
  ``epsilon-finite basis of master integrals for the integration-by-parts method,''
  Nucl.\ Phys.\ B {\bf 742}, 208 (2006)
  [hep-ph/0601165].

\bibitem{Chetyrkin:2006bj} 
  K.~G.~Chetyrkin, M.~Faisst, J.~H.~Kuhn, P.~Maierhofer and C.~Sturm,
  ``Four-Loop QCD Corrections to the Rho Parameter,''
  Phys.\ Rev.\ Lett.\  {\bf 97}, 102003 (2006)
  [hep-ph/0605201].

\bibitem{Boughezal:2006xk} 
  R.~Boughezal and M.~Czakon,
  ``Single scale tadpoles and O(G(F m(t)**2 alpha(s)**3)) corrections to the rho parameter,''
  Nucl.\ Phys.\ B {\bf 755}, 221 (2006)
  [hep-ph/0606232].

\bibitem{Bejdakic:2006vg} 
  E.~Bejdakic and Y.~Schroder,
  ``Hypergeometric representation of a four-loop vacuum bubble,''
  Nucl.\ Phys.\ Proc.\ Suppl.\  {\bf 160}, 155 (2006)
  [hep-ph/0607006].

\bibitem{Faisst:2006sr} 
  M.~Faisst, P.~Maierhoefer and C.~Sturm,
  ``Standard and epsilon-finite Master Integrals for the rho-Parameter,''
  Nucl.\ Phys.\ B {\bf 766}, 246 (2007)
  [hep-ph/0611244].

\bibitem{Bejdakic:2009zz} 
  E.~Bejdakic,
  ``Feynman integrals, hypergeometric functions and nested sums,''
  Doctoral thesis, Bielefeld Univ. October 2009. 
  \begin{verbatim}http://pub.uni-bielefeld.de/publication/2304683\end{verbatim}
  
\bibitem{Lee:2010hs} 
  R.~N.~Lee and I.~S.~Terekhov,
  ``Application of the DRA method to the calculation of the four-loop QED-type tadpoles,''
  JHEP {\bf 1101}, 068 (2011)
  [1010.6117].

\bibitem{Liu:2015fxa} 
  T.~Liu and M.~Steinhauser,
  ``Decoupling of heavy quarks at four loops and effective Higgs-fermion coupling,''
  Phys.\ Lett.\ B {\bf 746}, 330 (2015)
  [1502.04719].

\bibitem{IBP}
K.~G.~Chetyrkin and F.~V.~Tkachov,
  ``Integration by Parts: The Algorithm to Calculate beta Functions in 4 Loops,''
  Nucl.\ Phys.\ B {\bf 192}, 159 (1981),
F.~V.~Tkachov,
  ``A Theorem on Analytical Calculability of Four Loop Renormalization Group Functions,''
  Phys.\ Lett.\ B {\bf 100}, 65 (1981).

\bibitem{Laporta:2001dd} 
  S.~Laporta,
  ``High precision calculation of multiloop Feynman integrals by difference equations,''
  Int.\ J.\ Mod.\ Phys.\ A {\bf 15}, 5087 (2000)
  [hep-ph/0102033].


\bibitem{MVI}
  M.~E.~Machacek and M.~T.~Vaughn,
  ``Two Loop Renormalization Group Equations in a General Quantum Field Theory. 1. Wave Function Renormalization,''
  Nucl.\ Phys.\ B {\bf 222}, 83 (1983).

\bibitem{MVII} 
  M.~E.~Machacek and M.~T.~Vaughn,
  ``Two Loop Renormalization Group Equations in a General Quantum Field Theory. 
  2. Yukawa Couplings,''
  Nucl.\ Phys.\ B {\bf 236}, 221 (1984).

\bibitem{Jack:1984vj} 
  I.~Jack and H.~Osborn,
  ``General Background Field Calculations With Fermion Fields,''
  Nucl.\ Phys.\ B {\bf 249}, 472 (1985).

\bibitem{MVIII} 
  M.~E.~Machacek and M.~T.~Vaughn,
  ``Two Loop Renormalization Group Equations in a General Quantum Field Theory. 
  3. Scalar Quartic Couplings,''
  Nucl.\ Phys.\ B {\bf 249}, 70 (1985).

\bibitem{Chetyrkin:2012rz} 
  K.~G.~Chetyrkin and M.~F.~Zoller,
  ``Three-loop $\beta$-functions for top-Yukawa and the Higgs 
  self-interaction in the Standard Model,''
  JHEP {\bf 1206}, 033 (2012)
  [1205.2892].

\bibitem{Chetyrkin:2013wya} 
  K.~G.~Chetyrkin and M.~F.~Zoller,
  ``$\beta$-function for the Higgs self-interaction in the Standard Model at three-loop level,''
  JHEP {\bf 1304}, 091 (2013)
  [1303.2890].

\bibitem{Bednyakov:2013eba} 
  A.~V.~Bednyakov, A.~F.~Pikelner and V.~N.~Velizhanin,
  ``Higgs self-coupling beta-function in the Standard Model at three loops,''
  Nucl.\ Phys.\ B {\bf 875}, 552 (2013)
  [1303.4364].

\bibitem{vanRitbergen:1997va} 
  T.~van Ritbergen, J.~A.~M.~Vermaseren and S.~A.~Larin,
  ``The Four loop beta function in quantum chromodynamics,''
  Phys.\ Lett.\ B {\bf 400}, 379 (1997)
  [hep-ph/9701390].

\bibitem{Czakon:2004bu} 
  M.~Czakon,
  ``The Four-loop QCD beta-function and anomalous dimensions,''
  Nucl.\ Phys.\ B {\bf 710}, 485 (2005)
  [hep-ph/0411261].
  
\bibitem{Chetyrkin:1997dh} 
  K.~G.~Chetyrkin,
  ``Quark mass anomalous dimension to O (alpha-s**4),''
  Phys.\ Lett.\ B {\bf 404}, 161 (1997)
  [hep-ph/9703278].

\bibitem{Vermaseren:1997fq} 
  J.~A.~M.~Vermaseren, S.~A.~Larin and T.~van Ritbergen,
  ``The four loop quark mass anomalous dimension and the invariant quark mass,''
  Phys.\ Lett.\ B {\bf 405}, 327 (1997)
  [hep-ph/9703284].

\bibitem{Baikov:2014qja} 
  P.~A.~Baikov, K.~G.~Chetyrkin and J.~H.~KŸhn,
  ``Quark Mass and Field Anomalous Dimensions to ${\cal O}(\alpha_s^5)$,''
  JHEP {\bf 1410}, 76 (2014)
  [1402.6611].
  
\bibitem{Marquard:2015qpa} 
  P.~Marquard, A.~V.~Smirnov, V.~A.~Smirnov and M.~Steinhauser,
  ``Quark Mass Relations to Four-Loop Order in Perturbative QCD,''
  Phys.\ Rev.\ Lett.\  {\bf 114}, no. 14, 142002 (2015)
  [1502.01030].

\bibitem{Martin:2014cxa} 
  S.~P.~Martin and D.~G.~Robertson,
  ``Higgs boson mass in the Standard Model at two-loop order and beyond,''
  Phys.\ Rev.\ D {\bf 90}, no. 7, 073010 (2014)
  [1407.4336].

\bibitem{Martin:2015lxa} 
  S.~P.~Martin,
  ``Pole mass of the W boson at two-loop order in the pure $\overline {MS}$ scheme,''
  Phys.\ Rev.\ D {\bf 91}, no. 11, 114003 (2015)
  [1503.03782].

\bibitem{Martin:2015rea} 
  S.~P.~Martin,
  ``Z boson pole mass at two-loop order in the pure MS-bar scheme,''
  Phys.\ Rev.\ D {\bf 92}, 014026 (2015)
  [1505.04833].

\end{thebibliography}
\end{document}